\documentclass[12pt,a4paper,twoside]{article}
\usepackage{epsfig}
\usepackage{baltlat6}  
\usepackage{wrapfig}
\usepackage{dcolumn}
\begin{document}
%%%%%%%%%%%

 \ \
 \vspace{-0.5mm}

\setcounter{page}{1}
\vspace{-2mm}

\titlehead{Baltic Astronomy, vol.\ts xx, xxx--xxx, 200x.}

\titleb{IMPROVED PARAMETERS OF THE HYDROGEN--DEFICIENT BINARY STAR KS\,PER}

\begin{authorl}
\authorb{T\~onu Kipper}{1} and
\authorb{Valentina G. Klochkova}{2}
\end{authorl}

\begin{addressl}
\addressb{1}{Tartu Observatory, T\~oravere, 61602, Estonia; tk@aai.ee} 

\addressb{2}{Special Astrophysical Observatory RAS, Nizhnij Arkhyz, 369167,
             Russia; valenta@sao.ru}
\end{addressl}

%
%\submitb{Received ..., 2008}
%
\begin{abstract}

Using the high resolution spectra secured with the Nasmyth Echelle
Spectrograph NES of the 6 meter telescope we analysed the
hydrogen-deficient binary star KS\,Per. The atmospheric
parameters derived are: $T_{\rm eff}$\,=\,9500$\pm$300\,K,
$\log g$\,=\,2.0$\pm$0.5, and $\xi_{\rm t}$\,=\,9.5$\pm0.5$\,km\,s$^{-1}$.
The hydrogen deficiency is H/He\,=\,3$\cdot10^{-5}$, iron abundance is
reduced by 0.8\,dex, nitrogen abundance is very high [N/Fe]\,=\,1.4, but
carbon and oxygen abundances are low. The star luminosity is $\log
L/L_{\odot}$\,=\,3.3. A complex absorption and emission structure of the
Na\,I D doublet was revealed. We suggest that the emission component forms
in the circumbinary gaseous envelope.
\end{abstract}


\resthead{Parameters of KS\,Per}{T.\,Kipper, V.\,Klochkova }

{
\sectionb{1}{INTRODUCTION}

Studies of the hydrogen-deficient binaries are interesting as these
objects could be the progenitors of type Ia (Iben \& Tutukov, 1993;
Parthasarathy et al. 2007) or type Ib (Uomoto 1986) supernovae. Only four
hydrogen-poor close binaries (HdBs) are currently known, which indicates
that these objects are very rare. The primary components in HdBs are
hydrogen-poor low-gravity A--F stars transferring matter to a secondary
star. The high luminosity is provided by their helium-burning shell around
a CO core. Nature of more massive but less luminous secondaries is unclear
since their spectra are detected only in far UV spectral region. The
hydrogen-deficiency of HdBs is believed to be due to case BB mass transfer
in a binary system where the primary loses mass for a second time as it
evolves through He burning in shell (Delgado \& Thomas 1981). 
The most famous star among HdBs is $\upsilon$\,Sagittarii. That star was
the first one for which the spectra of both components were seen by Dudley
\& Jeffery (1990) who deconvolved the IUE 115--320\,nm spectra. They
determined the orbits, the mass ratio and the minimum mass for the
primary. Their results of the minimum mass indicate that $\upsilon$\,Sgr
could be a progenitor of SN\,Ib.

A semi-regular pulsating star KS\,Per($\equiv$HD\,30353 $\equiv$Bidelman's
star) was also found to be markedly deficient in hydrogen and to be a
spectroscopic binary (Bidelman, 1950), and so the member of this small
group. The pulsations which cause the semiregular light variations with a
period of about 30 days were found by Osawa, Nishimura \& Nariai (1963).

KS\,Per is an optical counterpart of an IR-source IRAS\,04453+4311 (its
IRAS--fluxes are given in Table\,1). Dudley \& Jeffery (1993) modelled the
star's infrared flux and estimated the effective temperature of its
circumstellar dust $T_{\rm eff}=1100$\,K. Presence of hot circumstellar
dust indicates recent or current mass loss. Indeed, Parthasarathy et al.
(1990) analysed UV spectra (IUE) of KS\,Per and found shortward-shifted
stellar wind profiles of various species. The terminal velocity from N\,V,
C\,IV and Si\,IV lines reaches $-$650\,km\,s$^{-1}$. The Mg\,II doublet
has P\,Cygni profile with $v_{\rm term}$\,=\,$-$415\,km\,s$^{-1}$. Other
lines show shifts about $-$200\,km\,s$^{-1}$. The large blueshifts are
presented also in optical region by the H$\alpha$ line in our spectra
(Fig.\,1).

Margoni et al. (1988) studied radial velocities of KS\,Per and combining
their own data with the earlier published ones they determined the orbital
period of 362.8\,d, semiamplitude $K$\,=\,48$\pm$2\,km\,s$^{-1}$, and mass
function $F(m)$\,=\,3.6$\pm$0.4\,${\mathcal M}_{\odot}$. Although the mass
function is relatively high, they did not find any trace of the companion
spectrum in the optical region.

The presence of highly ionized species (N\,V and C\,IV) together with the
UV-excess in the flux shortward of 180\,nm~ suggest that the companion is
an early B-type star. The mass-function $F(m)$ values derived from the
single-lined spectroscopic orbits suggest that the primaries of
hydrogen-poor close binaries may have CO core masses about
1--1.3\,${\mathcal M}_{\odot}$ with extended outer envelopes. They may be
in the post-AGB phase (Parthasarathy et al. 2007). The secondaries may have
masses 3-4\,${\mathcal M}_{\odot}$. From the early observations of KS\,Per
no radiation from the secondary was found. This led Zeldovich \& Guseinov
(1966) to suppose that it is a collapsed star.

Whether KS\,Per is a supergiant, as the spectroscopic criteria show, is not
definitely known (Bidelman 1950). Bidelman found from the interstellar
reddening $E(B-V)=0.44$ the distance about 2\,kpc which means that the star
is rather luminous supergiant ($M_V\approx-5^m$).

Most of the hydrogen deficient stars are rich in carbon (see, for example,
Kipper \& Klochkova 2005, 2006), but KS\,Per is not despite its spectrum
resembles that of R\,CrB stars. In order to obtain accurate chemical
abundances we performed high resolution spectral observations covering
large spectral region. The results are reported in the following.

\sectionb{2}{SPECTRAL OBSERVATIONS  AND REDUCTION}

Our high resolution spectra (R\,=\,60000) were obtained with the Nasmyth
Echelle Spectrograph NES (Panchuk et al. 2007) permanently located at the
Nasmyth focus of Russian 6\,meter telescope. The spectrograph was equipped
with an image slicer (Panchuk et al. 2007). As a detector a CCD camera
with $2048\times2048$ pixels ($15\times15\mu$m, readout noise $7e^-$)
produced by the Copenhagen University Observatory was used. The spectral
regions 521--668\,nm were registered on August, 31 2007 (JD\,2454344) and
394--540\,nm on September, 01 2007 (JD\,2454345). The spectra cover
394-668\,nm without gaps until 610\,nm. To increase the signal to noise
ratio 8 individual exposures with exposure time 1800\,sec were obtained. 
                                                                                
The spectra were reduced using the NOAO astronomical data analysis
facility IRAF. The use of image slicer results in three parallel strips of
spectra in each order. These strips are wavelength shifted. Therefore all
strips were reduced separately, linearized in the wavelength and coadded.
We checked the accuracy of this procedure (Kipper \& Klochkova 2005) and
found that the wavelengths of the terrestial lines in the stellar spectrum
were reproduced within a few 0.001\,\AA-s.

The spectra were radial velocity corrected due to the solar motion using
IRAF task ``rvcorrect''. After that all the spectra were coadded. The
continuum was placed by fitting low order spline functions through the
manually indicated points in every order.

\sectionb{3}{SPECTRAL ANALYSIS}

\subsectionb{3.1}{Radial velocity}

For spectroscopic binaries the radial velocity is an important parameter.
Based on our high $S/N$ and high resolution spectra covering large
wavelength interval the accurate radial velocity was measured for the
observed orbital phase. First, the 8 lines listed by Margoni et al. (1988)
were identified. From those the H$\delta$ line and a Mg\,II doublet were
excluded due to serious blending. The remaining 6 lines provide a
provisional radial velocity $-31.9\pm1.4$\,km\,s$^{-1}$. With this
velocity the other lines were identified. Due to low continuous opacity
the number of lines in KS\,Per spectrum is much larger than in normal
stellar spectra making the blending of serious issue. Using the most
symmetrical and presumably unblended Fe\,II lines (32 lines) the
heliocentric radial velocity $v_{\odot}$\,=\,$-32.3\pm1.7$\,km\,s$^{-1}$
was determined. {No dependence of Vr on line formation depth was found.}
To get the velocity relative to LSR the correcion $-$5.08\,km\,s$^{-1}$
should be added.

The systemic heliocentric radial velocity of about +3\,km\,s$^{-1}$ was
found by Margoni et al. (1988). That velocity agrees with the pattern of
Galactic plane rotation in direction of KS\,Per (Dame et al. 2001)
for distances less than 1\,kpc.

In Fig.\,1 the very strong Na\,I\,D doublet is plotted in velocity scale.
Danziger et al. (1967) have used these lines for finding the interstellar
(IS) reddening towards KS\,Per assuming that the radial velocity of the IS
line is about +5\,km\,s$^{-1}$. This is clearly not the case as in Fig.\,1
the lines are blueshifted by more than 20\,km\,s$^{-1}$. In Bidelman's paper
(1950) these lines were found to be  redshifted by about 25\,km\,s$^{-1}$. This radial
velocity variation shows that these lines are of circumstellar origin.
Note the P\,Cygni profile of red wings. Emission components of the
Na\,I\,D lines are at the nearly zero radial velocity which agrees with
the systemic velocity. Therefore we propose that this emission forms 
in the circumbinary gaseous envelope.

\subsectionb{3.2}{Model atmospheres}

Early estimates of chemical composition of KS\,Per by Wallerstein et
al. (1967) and Nariai (1967) showed that the atmosphere is very
hydrogen-poor H/He$\approx10^{-4}$ (by number). The abundance of carbon
compared to that of nitrogen is also low. Logarithmic abundances by
numbers found by Wallerstein et al. (1967) were He=11.6, C=6.2, and N=9.2.
Dudley \& Jeffery (1993) have also found that carbon abundance in HdBs is
lower than in extreme helium stars.

Such hydrogen-deficient atmospheric models were computed in Armagh
Observatory using the code STERNE (Jeffery et al. 2001). These models take
into account both hydrogen-deficiency and metal-line blanketing. From
Armagh Observatory database the He+N models grid $h00he99n003$ was chosen
for the present analysis  (see the Web--address {\it http://star.arm.ac.uk}).

\subsectionb{3.3}{Atmospheric parameters}

Spectral type of KS\,Per is A5Iap (SIMBAD database). For normal stars this
would correspond to $T_{\rm eff}$\,=\,8600\,K and $\log g$\,=\,2.0. But, as
Danziger et al. (1967) have noted, when the lines are greatly enhanced by
low opacity, the approximate spectral type cannot be used to infer the
effective temperature or absolute magnitude with high accuracy. Early
estimate by Nariai (1963) using photometry and line blanketing data is
$T_{\rm eff}$\,=\,8400\,K. The same autor (Nariai 1967) found using
hydrogen-deficient model atmospheres $T_{\rm eff}$\,=\,11000$\pm$1000\,K,
$\log g$\,=\,1$\pm$1, and $\xi_{\rm t}$\,=\,18\,km\,s$^{-1}$. At the same
time Wallerstein et al. (1967) also analysed the star and found ionization
temperature $T_{\rm ion}$\,=\,10080\,K. They let $T_{\rm ion}$\,=\,$T_{\rm
eff}$. Danziger et al. (1967) used interstellar polarization and reddening
and obtained for KS\,Per $T_{\rm eff}$\,=\,10000\,K, $\log g$\,=\,2.0,
$E(B-V)$\,=\,0.35 and $M_V$=$-3.2^m$.

Dudley \& Jeffery (1993) analysed all four hydrogen-deficient close
binaries modelling their flux distribution. The model atmospheres were
calculated with the assumed abundances: $n_{\rm H}$\,=\,0.0002, $n_{\rm
He}$\,=\,0.998, $n_{\rm C}$\,=\,0.00007, and $n_{\rm N}$\,=\,0.02. For
KS\,Per they obtained parameters: $T_{\rm eff}$\,=\,12500$\pm$500\,K,
$E_{B-V}$\,=\,0.55$\pm$0.10, and $d$\,=\,3.9\,kpc. Recently Pan\-dey (2006)
estimated $T_{\rm eff}$\,=\,10500\,K, $\log g=1.5$, and $\xi_{\rm
t}$\,=\,10\,km\,s$^{-1}$.

We started with this last estimate. Forcing the excitation and ionization
equilibriae of Fe\,I and Fe\,II we ended up with $T_{\rm
eff}$\,=\,9500$\pm300$\,K and $\log g$\,=\,2.0$\pm0.5$. For this procedure
we were forced to extrapolate Armagh atmospheric models towards lower
temperatures by 1000\,K. For the microturbulent parameter we got somewhat
different values for Fe\,I and Fe\,II 8.3 and 9.4\,km\,s$^{-1}$
respectively. Other elements lines were best fitted with similar
microturbulent velocity values. We adopted the weighted by the number of
used lines mean microturbulent velocity $\xi_{\rm
t}=9.5\pm0.5$\,km\,s$^{-1}$. This is considerably lower value than
18\,km\,s$^{-1}$ found by Nariai (1967). For $\upsilon$\,Sgr Leushin
(2001) found $\xi_{\rm t}=8\div12$\,km\,s$^{-1}$ depending on spectral
region.

\subsectionb{3.4}{Hydrogen content}

The hydrogen to helium ratio was estimated by Wallerstein et al. (1967) to
be near 1$\cdot10^{-4}$ and by Nariai (1964) 3$\cdot10^{-4}$. Using the
new Armagh hydrogen-deficient model atmospheres we synthesized the
H$\beta$, H$\gamma$ and H$\delta$ lines (the H$\beta$ and H$\delta$ lines
are shown in Fig.\,4). According to these figures H/He is close to
$2\cdot10^{-5}$. Using the fitting of measured equivalent widths we got
from the H$\gamma$ and H$\delta$ lines H/He\,=\,3$\cdot10^{-5}$, but from
the H$\beta$ line much less 2$\cdot10^{-6}$. This indicates that the
H$\beta$ line is seriously filled with emission (see left panel of
Fig.\,4). Also the the red wing of the H$\beta$ line is also slightly in
emission showing P\,Cygni like profile. The central parts of the H$\gamma$
and H$\delta$ lines are also filled with emission as could be judged from
Fig.\,4 where the synthesized relatively deep Doppler cores were not
observed. Note that the same peculiarities of the Balmer lines profiles
were observed in the spectrum of the another HdB star $\upsilon$\,Sgr
(Leushin 2001).

\subsectionb{3.5}{Abundances}
 
The abundances of other elements were derived with the help of Kurucz's
program WIDTH5 together with the Armagh hydrogen-deficient model
atmospheres. The sources of oscillator strengths are indicated in the
Table~2, where the results are listed. Measured equivalent widths of
lines, used oscillator strengths and derived abundances are listed in
Table~4. For most lines the oscillator strengths by Thevenin (1989, 1990)
and for C, N and O the data by Wiese et al. (1996) were used. 

When the hydrogen is no longer the most abundant element, the usual scale,
where the logarithmic abundance of hydrogen is 12.00, is no more convenient.
In WIDTH5 the abundances are determined relative to total number of atoms.
In the extremely hydrogen-poor case this is the number of He atoms which
in logarithmic scale should be taken 11.54 if one wishes to normalize
$\log \Sigma \mu_i\varepsilon(i)=12.15$ as in the solar case.
  
The low hydrogen abundance and consequently low continuous opacity leads to
very large number of lines in stars spectrum. This in turn makes difficult
to find nonblended lines for abundance determinations and increases the
errors. The errors indicated in Table~2 are due to differing results from
different lines. The systematic errors due to errors in $T_{\rm eff}$,
$\log g$ and $\xi_{\rm t}$, found by changing these parameters, are less
than these indicated errors.
										
First of all, the results show that KS\,Per is metal poor with the mass
fraction of Fe reduced by 0.8\,dex in accord with Nariai (1967) finding
that the abundances of metals are about $10^{-2}\div10^{-3}$ times that of
$\alpha$\,Cyg. The abundance of He, checked using 5 He\,I lines, is
$11.54\pm0.23$. No C\,I or C\,II lines were reliably identified. Only a
C\,I line at 477.174\,nm, which in the solar spectrum is blended
with a Fe\,I line, was measured providing C abundance of 7.5. This is less
than the input abundance.
The He abundance also provides a circumstancial evidence for low carbon
abundance. If the input C abundance is raised by 0.3\,dex, the needed He
abundance should be raised by 0.1\,dex. The oxygen abundance is also low.
Nitrogen is enhanced by large amount indicating that the surface
material is primarily CNO processed. The $s-$process produced elements
seem to be enhanced but just at the amount not exceeding the determination
errors which are comparatively large due to small number of used lines and
heavy blending.

Recall that the star studied is an counterpart of an IR-source. For such
an object with a dusty-gaseous circumstellar envelope one could expect
that the chemical abundances pattern is modified by selective separation
processes. For example, a part of Fe deficit could be caused by gas-dust
separation. The position of the star very close to the Galactic plane and
its proper motion  permit us to propose that it belongs to Pop.\,I and has
normal metallicity.
The CNO--triad, zink and sulfur are unaffected by selective fractioning
processes. Since the abundances of CNO may vary due to nuclear reactions
in the course of the star's evolution, the behaviour of Zn and S are
critical for finding the efficiency of selective depletion. According to
Wheller et al. (1989) and Timmes et al. (1995) the Zn abundance varies
together with the Fe abundance over a wide range of metallicity and could
be used as criterion of initial star metallicity.

Unfortunately, when studying the abundance of Zn in KS\,Per we encountered
the difficulties which we were not able to overcome. Two stronger Zn\,I
lines at 481.054 and 636.235\,nm were found at the wavelengths 481.010 and
636.165\,nm corresponding to radial velocities -32.9 and
-33.0\,km\,s$^{-1}$. The measured equivalent widths 23.0 and 22.3\,pm,
however, correspond to abundances of 5.97 and 6.22 in the scale of Table
2, when the oscillator strengths by Bi\'emont \& Godefroid (1980) were
used. Solar abundance of zinc is about 4.6. Therefore we supposed that
these lines do not belong to Zn\,I but we were not able to identify their
origin.

We searched also for a weaker Zn\,I line at 472.216\,nm and found a line
at 472.152\,nm with $W_\lambda$=1.3\,pm. This will give the abundance 4.32
of zinc. However, the corresponding radial velocity is
-40.6\,km\,s$^{-1}$, which is much too different from the mean found from
other lines.

The relative abundance of volatile element sulfur [S/Fe] in the atmosphere
of KS\,Per is within the errors close to the solar one. In order to verify
additionally the possible depletion pattern we plotted relative abundances
[El/Fe] from Table\,2 versus condensation temperature $T_{cond}$ (Lodders
2003) (Fig.\,5). No significant relation between $T_{cond}$ and abundances
was found. Thus we do not see any trace of selective separation in the
circumstellar envelope of KS\,Per.

In the other HdB $\upsilon$\,Sgr iron was found to be slightly
overabundant (Leushin 2001). But the position of $\upsilon$\,Sgr in the
Galaxy is completely different from KS\,Per.

\subsectionb{3.6}{Luminosity}

Bidelman (1950) in his pioneering work estimated using spectroscopic
criteria $M_V\approx-5^m$, $E(B-V)=0.4$ and distance about 2\,kpc for
KS\,Per. But, as already noted, for greatly enhanced lines these estimates
could have large errors. Danziger et al. (1967) estimated from
polarization measurements and comparisons with neighboring stars
0.25$<E(B-V)<$0.45 and distance modules 9.6$\div$11$^m$. They found the
same numbers from Na\,I\,D IS line intensities, but as it turned out, the
Na\,I\,D doublet in KS\,Per spectrum is entirely of circumstellar (CS)
origin (see sec.3.1). The absence of IS lines is in accord with KS\,Per
position in so called Auriga Gap with very few molecular clouds. In
microwave frequencies there are no objects towards KS\,Per (Jardine 2008).
Parthasarathy et al. (1990) estimated from width-luminosity relationship
of Mg\,II~k emission line $M_V$\,=\,$-4\div-6^m$, and 0.30$<E(B-V)<$0.45
in according with Danziger et al. (1967). This means that
$E(B-V)$$\approx$0.35 is fairly well established.

The systemic heliocentric radial velocity of KS\,Per is about
+3\,km\,s${-1}$ (Margoni et al. 1988) or $v_{\rm LSR}=0$. This means that
the distance to KS\,Per is probably less than 1\,kpc as the velocities due
to galactic rotation in that direction are negative. According to Dame et
al. (2001) the CO emission in direction of KS\,Per has a zero velocity
component corresponding to local gas.

According to numerical code for galactic extinction by Hakkila et al.
(1997) $A_V$ reaches $1.5\pm0.4$ already at the distance 1\,kpc. With
$R\sim3.1$ this corresponds to $E(B-V)=0.48$ indicating that the
distance should be less than 1\,kpc. Therefore we end up with the
estimate $M_V$$\approx-3.3^m$ and $\log L/L_{\odot}$\,$\approx 3.3$. 

\sectionb {5}{CONCLUSION}

We have found that the chemical composition of metal- and
hydro\-gen-deficient pulsating star KS\,Per corresponds to the material
which is primarily CNO processed without indications of triple$-\alpha$
processing. During the CNO processing nitrogen is enhanced at the expense
of carbon and oxygen. This happens during the first and second dredge-up
and one could not decide whether the star has passed the second dredge-up.
Soon after that the star will burn He sporadically via the triple$-\alpha$
process. The low carbon abundance shows that the donor star had not yet
experienced the third dredge-up. In this sense it differs from the other
group of hydrogen-poor stars -- extreme helium stars (EHes) in which the
the carbon abundance is considerably enhanced.

Abundances in KS\,Per are quite close to the ones in its sibling
$\upsilon$\,Sgr except the abundance of neon and iron. In KS\,Per the Ne
abundance is nearly normal but in $\upsilon$\,Sgr it is greatly enhanced
(Leushin et al. 1998). This indicates that KS\,Per is in somewhat earlier
evolutionary phase than $\upsilon$\,Sgr.

In Table 3 we compare the abundances in KS\,Per with the ones in a cool
extreme helium star LS\,IV-14$^{\rm o}$109 with close atmospheric
parameters ($T_{\rm eff}=9500$\,K, $\log g=0.9$ and Fe abundance) (Pandey
et al. 2001)). Chemical abundances pattern of KS\,Per as a whole, with the
exception of CNO-group, resembles that of EHes.
Excess of heavy metalls (Y, Zr and Ba) is, however, not statiscally
significant. Note also very high abundance of Ne in LS\,IV-14$^{\rm o}$109
compared to KS\,Per  at close metallicity of both stars.

Opposite to a pair of HdBs KS\,Per and $\upsilon$\,Sgr, EHes are not
binary systems and do not show the infrared excess. According to one
popular scenario EHes could be produced by merging of a He white dwarf
with a CO white dwarf. HdBs cannot be regarded as precursors of EHes as
the mass of the components would be above the Chandrasekhar limit and
rather a supernova event may result (Morrison 1988).

We revealed a complex emission\,+\,absorption profile of the Na\,I\,D and
suggest that the emission component forms in the circumbinary gaseous
envelope. To obtain more realistic interpretation of the Na\,I\,D doublet
profile one needs both spectral monitoring and spectropolarimetry with
high spectral resolution. Based on photoelectric polarimetric
observations, Pfeiffer \& Koch (1973) reported that KS\,Per possess
significant and variable degree of polarization, $P$\,=\,$1.84\div2.18\%$.

We have found that the luminosity of KS\,Per is $\log L/L_{\odot}\approx3.3$ 
which together with $T_{\rm eff}=9500$\,K corresponds to spectral type A2II.
This is somewhat different from the result of direct spectral
classification A5Iap (SIMBAD data base). Again, compared to
$\upsilon$\,Sgr with $M_V=-4.8\pm1.0$ (Kameswara Rao \& Venugopal 1985)
KS\,Per is less luminous.

\vskip 1.5cm ACKNOWLEDGEMENTS.\ This research was supported by the
Estonian Science Foundation grant nr.~6810 (T.\,Kipper).
V.\,Klochkova acknowledges support by the Russian Foundation for Basic
Research (project 08-02-00072\,a), the fundamental research program
``Extended Objects in the Universe'' of the Division of Physical Sciences
of the Russian Academy of Sciences, and the program ``Origin and Evolution
of Stars and Galaxies'' of the Presidium of the Russian Academy of
Sciences.

This research has made use of the SIMBAD database, operated at CDS,
Strasbourgh France.

\References{}
\refb
Artru M.C., Jamar C., Petrini D. \& Praderie F. 1981, A\&A, {\bf 44}, 171
\refb
Asplund M., Gustafsson B., Lambert D.L. \&  Rao N.K. 2000, A\&A, {\bf 353}, 287
\refb
Asplund M., Grevesse N. \& Sauval A.J. 2005, ASP Conf. Ser., {\bf 336}, 25 
\refb
Bidelman W.P. 1950, ApJ, {\bf 111}, 333
\refb
Bi\'emont E. \& Godefroid M. 1980, A\&A, {\bf 84}, 361
\refb
Dame T.M., Hartmann D. \& Thaddeus P., 2001, ApJ, {\bf 547}, 792
\refb
Danziger I.J., Wallerstein G. \& B\"ohm-Vitense E. 1967, ApJ, {\bf 150}, 239
\refb
Delgado A.J. \& Thomas A.-C. 1981, A\&A, {\bf 96}, 142
\refb
Drilling J.S. \& Sch\"onberner D. 1982, A\&A, {\bf 113}, L22
\refb
Dudley R.E. \& Jeffery C.S. 1990, MNRAS, {\bf 247}, 400
\refb
Dudley R.E. \& Jeffery C.S. 1993, MNRAS, {\bf 262}, 945
\refb
Hakkila J., Myers J., Stidham B. \& Hartmann D. 1997, AJ, {\bf 114}, 2043
\refb
van Hoof P. 1999,  {\it http://www.pa.uky.edu/$\sim$peter/atomic}
\refb
Jardine K. 2008, {\it http://galaxymap.org/drupal/node/100}
\refb
Jeffery C.S., Woolf V.M. \& Pollaco D.L. 2001, A\&A, {\bf 376}, 497
\refb
Iben I.Jr. \& Tutukov A.V. 1993, ApJ, {\bf 418}, 343
\refb
Kameswara Rao N. \& Venugopal V.R. 1985, J. Ap.\,\&\,A., {\bf 6}, 101
\refb
Kipper T. \& Klochkova V.G. 2005, Baltic Astronomy, {\bf 14}, 215
\refb
Kipper T. \& Klochkova V.G. 2006, Baltic Astronomy, {\bf 15}, 531
\refb
Leushin V.V., Snezhko L.I. \& Chuvenkov V.V. 1998, Astron. Lett., {\bf 24}, 45
\refb
Leushin V.V. 2001, Astron. Lett., {\bf 27}, 643
\refb
Lodders K. 2003, ApJ, {\bf 591}, 1220
\refb
Margoni R., Stagni R. \& Mammano A. 1988, A\&AS, {\bf 75}, 157
\refb
Munari U. \& Zwitter T. 1997, A\&A, {\bf 318}, 269
\refb
Nariai K. 1963, PASJ, {\bf 15}, 449
\refb
Nariai K. 1967, PASJ, {\bf 19}, 63
\refb
Osawa K., Nishimura S. \& Nariai K. 1963, PASJ, {\bf 15}, 313
\refb
Panchuk V.E., Klochkova V.G., Najdenov I.D. \& Yushkin M.V.  2007,
        {\it In: ''UV Astronomy: Stars from birth to death''},
        eds. A.~I.~de~Castro \& M.~A~Barstow. Madrid,
	UCM Editorial Complutense, 179
\refb
Pandey G., Kameswara Rao N., Lambert D.L., Jeffery C.S. \& Asplund M. 2001,
             MNRAS, {\bf 324}, 937
\refb
Pandey G. 2006, ApJ, {\bf 648}, L143
\refb
Parthasarathy M., Hack M. \& Tektunali G. 1990, A\&A, {\bf 230}, 136
\refb
Parthasarathy M., Branch D., Jeffery D.J. \& Baron E. 2007,
astro-ph/0703415
\refb
Pfeiffer R.J. \& Koch R.H. 1973, IBVS, No.\,780
\refb
Thevenin F. 1989, A\&AS, {\bf 77}, 137
\refb
Thevenin F. 1990, A\&AS, {\bf 82}, 179
\refb
Timmes F.X., Woosley S.E. \& Weaver T. 1995, ApJS, {\bf 98}, 617
\refb
Uomoto A. 1986, ApJ, {\bf 310}, L35
\refb
Wallerstein G., Greene T.F. \& Tomley L.J. 1967, ApJ, {\bf 150}, 245
\refb
Wheeler J.C., Sneden C. \& Truran J.W. 1989, A\&A, {\bf 27}, 279
\refb
Wiese W.L., Fuhr J.R. \& Deters T.M. 1996, J. Chem. Ref. Data, Mono. 7
\refb
Zeldovich Ya.B. \& Guseinov O.H. 1966, ApJ, {\bf 144}, 840

\clearpage
\newpage
\begin{figure}[h!]
\vskip2mm
\centerline{{\psfig{figure=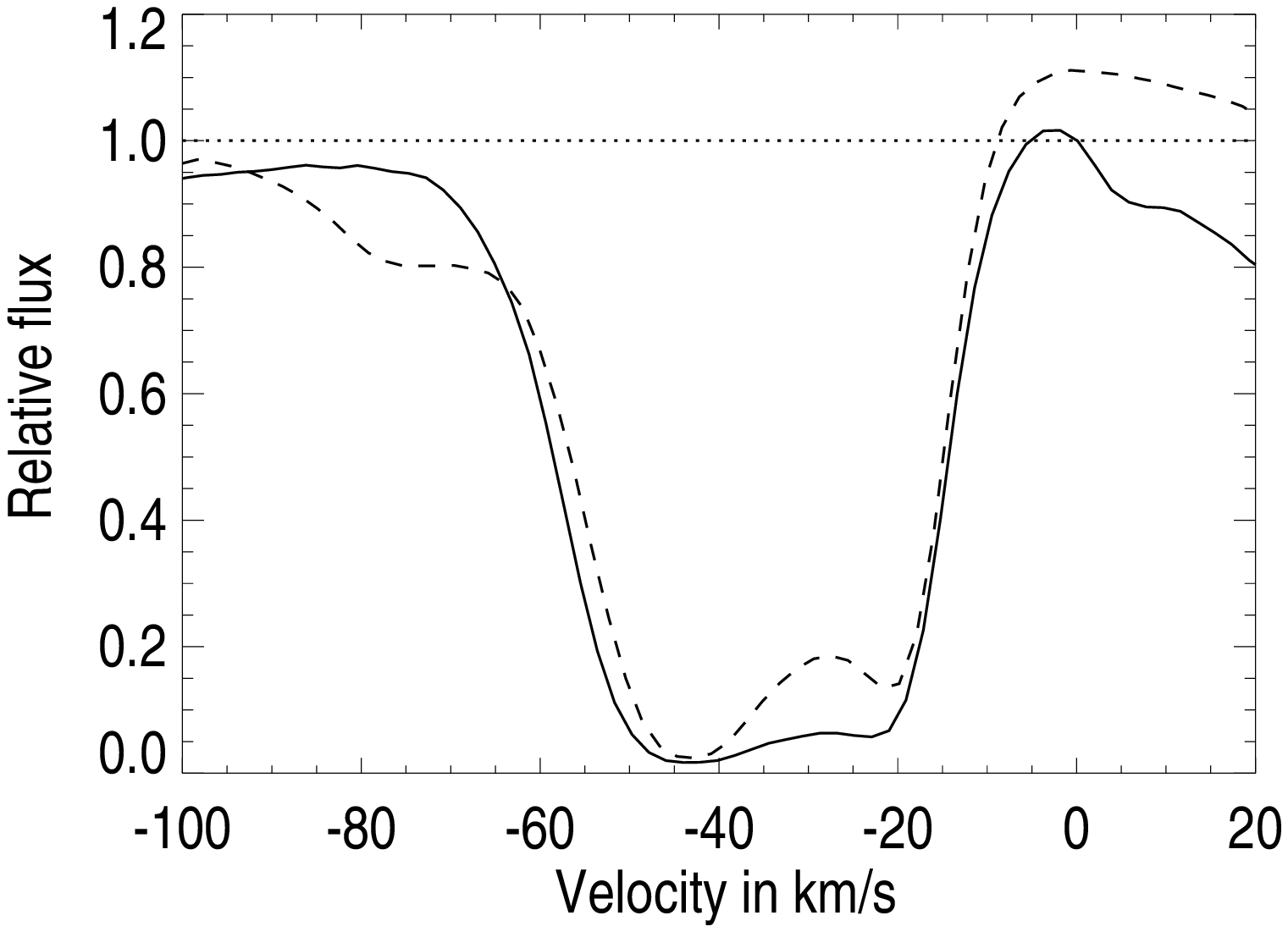,width=105truemm, bb=0 0 500 350,   angle=0,clip=}}
{\hspace{-0.3cm}\psfig{figure=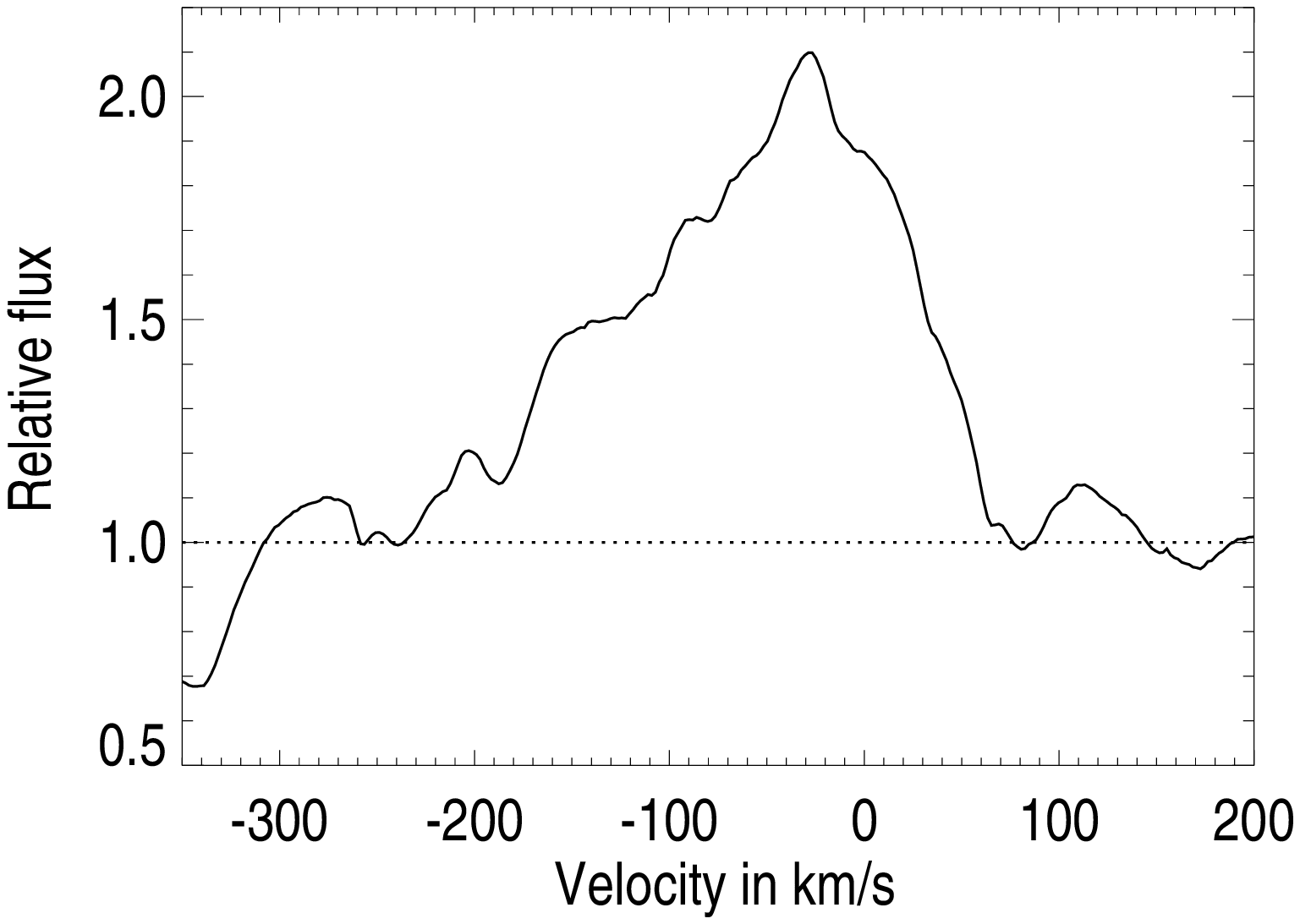,width=90truemm, bb=70 0 500 350,  angle=0,clip=}}}
\vskip2mm
\captionb{1}{The Na\,D doublet in the KS\,Per spectrum in velocity scale (left).
             The D$_2$ component is plotted with solid line, the D$_1$ with
	     dashed line. For both components the red half of the line show
	     P\,Cygni profiles, the blue component is also of CS origin.
	     In the right panel the H$\alpha$ emission is depicted in velocity scale.}
\vskip2mm
\end{figure}

\clearpage
\newpage
\begin{figure}[h!]
\vskip2mm
\centerline{{\psfig{figure=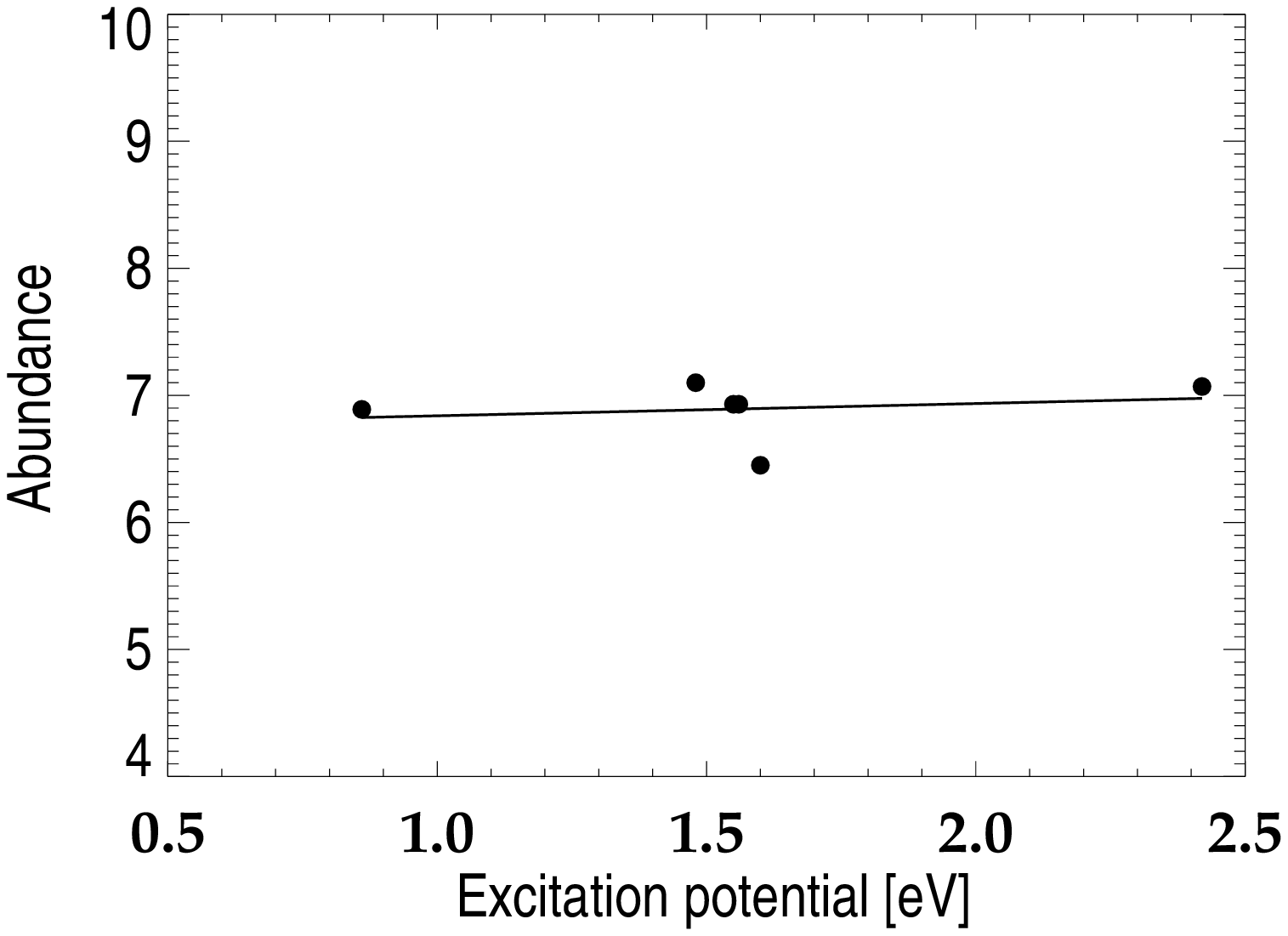,width=105truemm, bb= 0 0 500 350,  angle=0,clip=}}
            {\hspace{-0.3cm}\psfig{figure=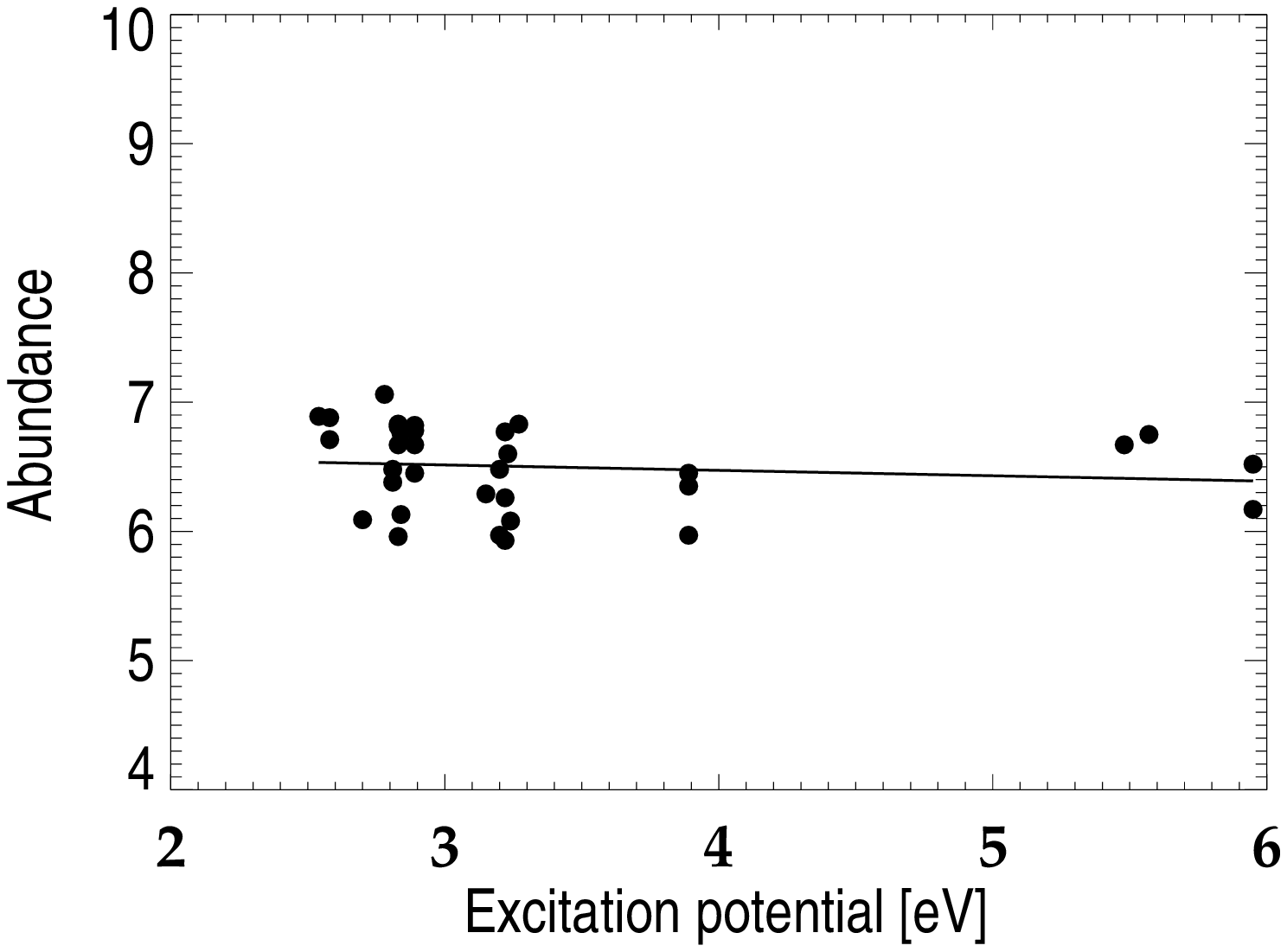,width=90truemm, bb= 70 0 500 350,  angle=0,clip=}}}
\vskip2mm
\captionb{2}{The dependence of the iron abundance on the lower excitation
             potential of the used lines for Fe\,I (left) and Fe\,II (right).
	     The hydrogen deficient model with $T_{\rm eff}=9500$\,K
             and $\log g=2.00$ together with  microturbulent velocities
	     8.3\,km\,s$^{-1}$ for Fe\,I and 9.4\,km\,s$^{-1}$ for
	     Fe\,II was used.}
\vskip2mm
\end{figure}

\clearpage
\newpage
\begin{figure}[h!]
\vskip2mm
\centerline{{\psfig{figure=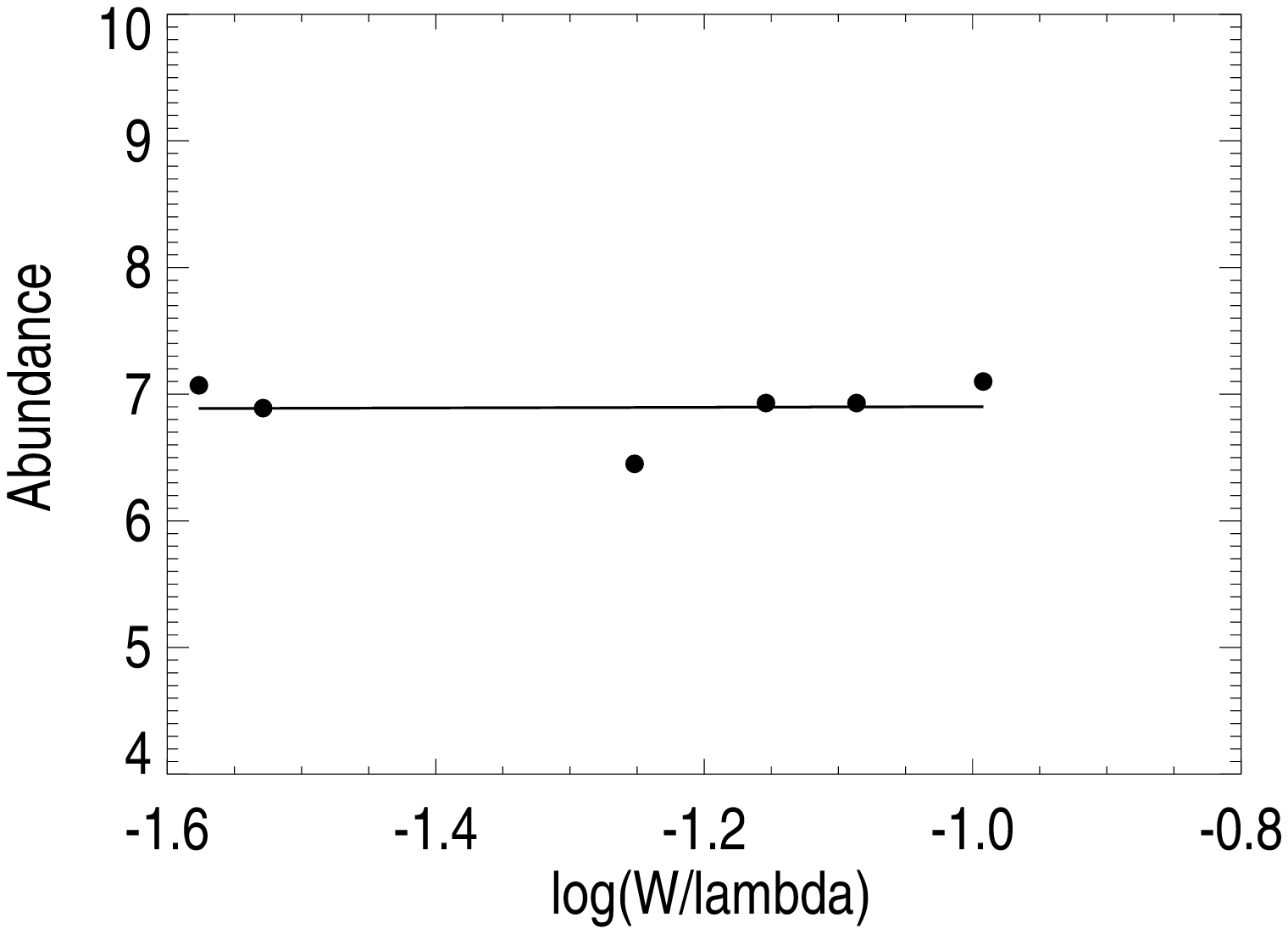,width=105truemm,bb=0 0 500 350, angle=0,clip=}}
           {\hspace{-0.3cm}\psfig{figure=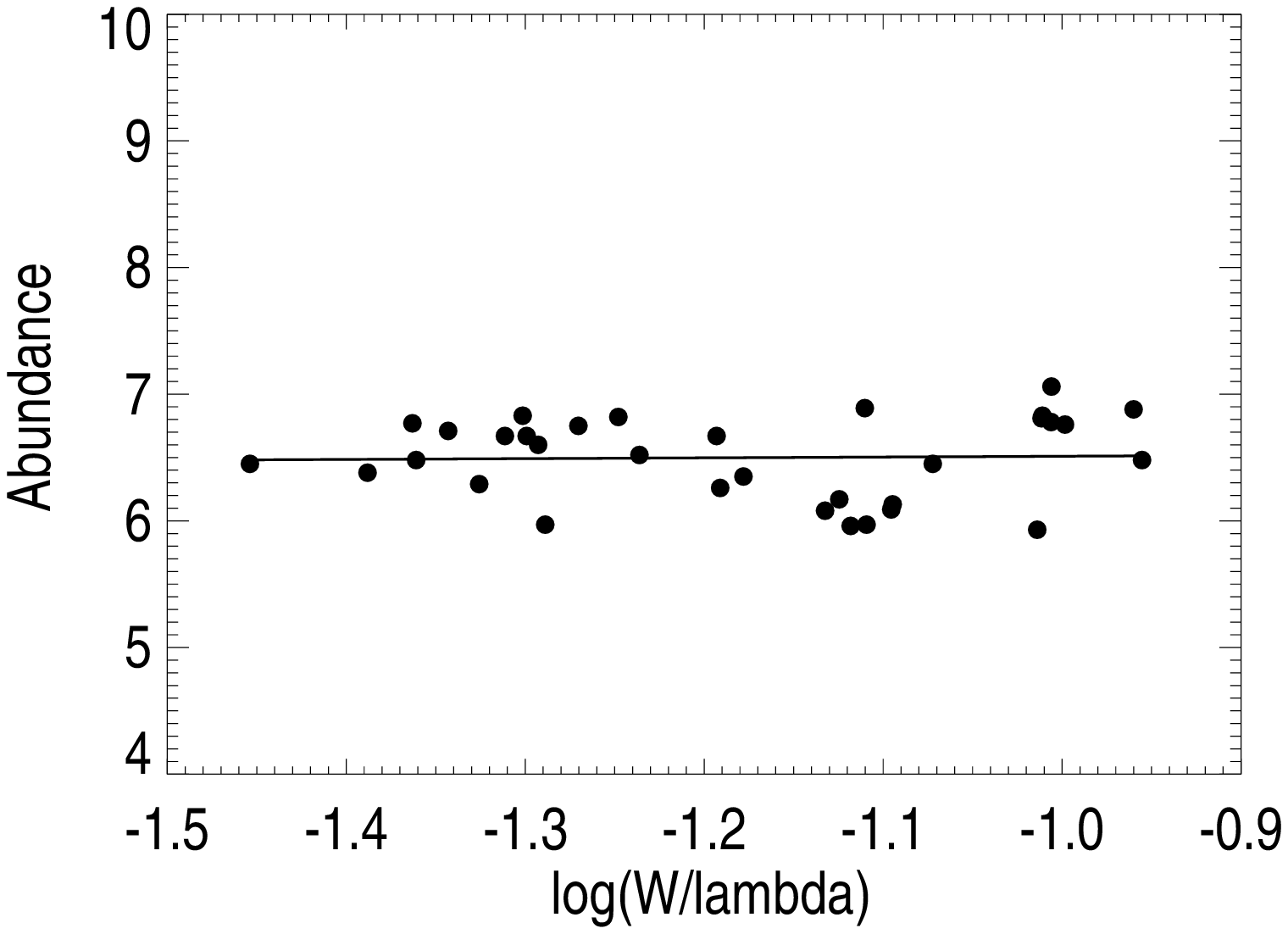,width=90truemm, bb=70 0 500 350, angle=0,clip=}}}
\vskip2mm
\captionb{3}{The dependence of the iron abundance on $W_{\lambda}/\lambda$
             of the used lines for Fe\,I (left) and Fe\,II (right). 
             The microturbulent velocities are 8.3\,km\,s$^{-1}$ for
             Fe\,I and 9.4\,km\,s$^{-1}$ for Fe\,II. }
\vskip2mm
\end{figure}

\clearpage
\newpage
\begin{figure}[h!]
\vskip2mm
\centerline{{\psfig{figure=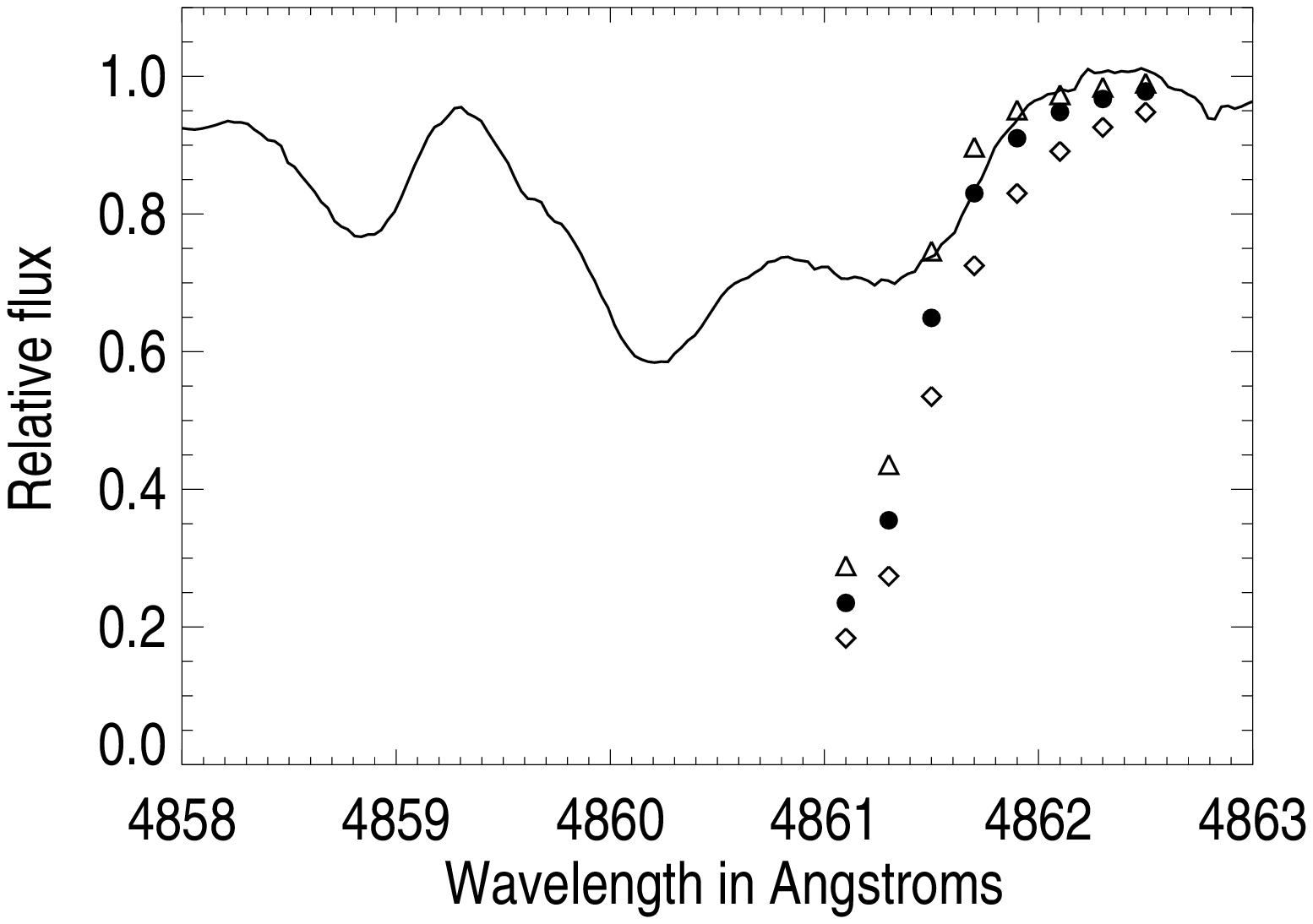,width=105truemm, bb=0 0 500 350,angle=0,clip=}}
           {\hspace{-0.3cm}\psfig{figure=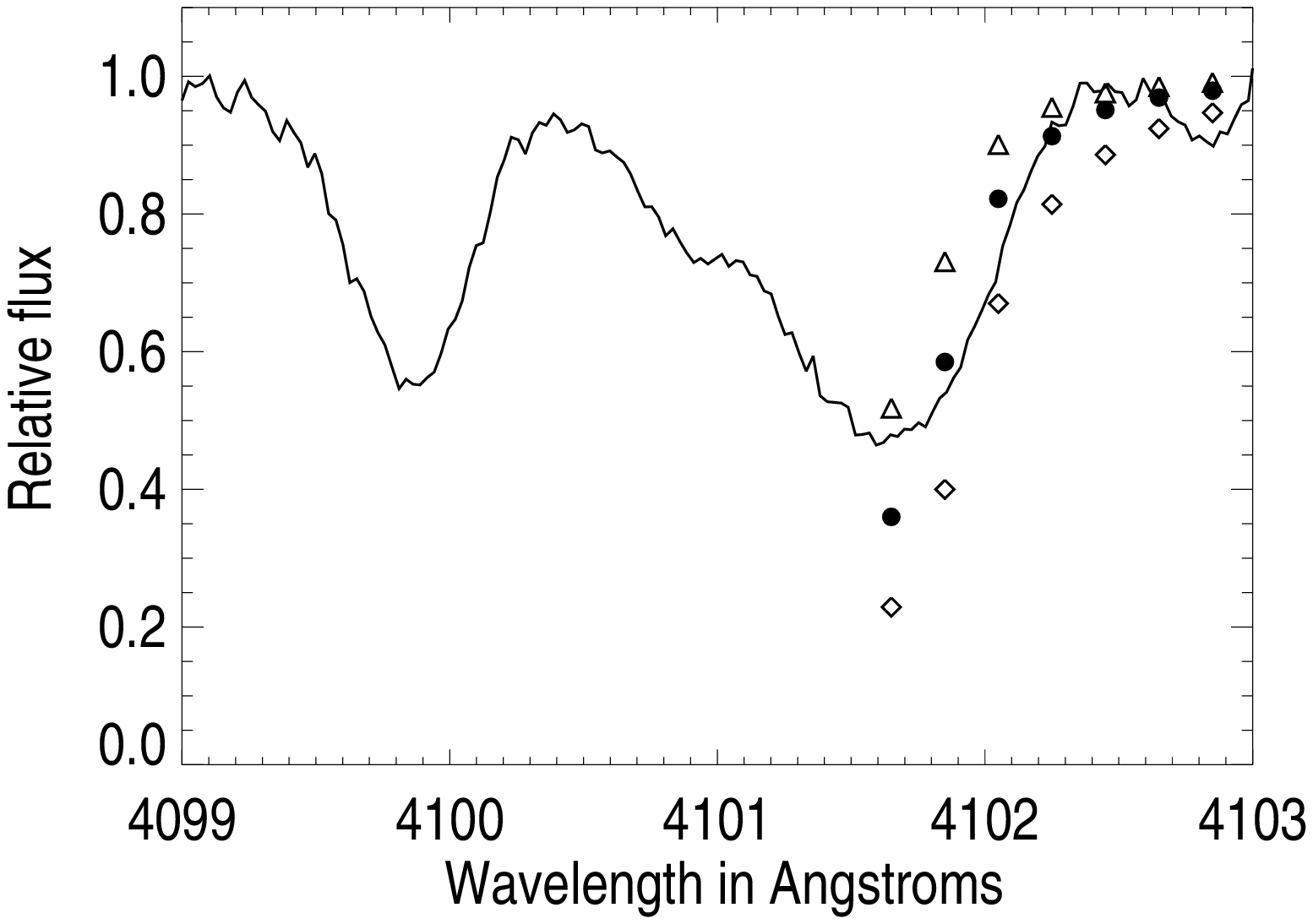,width=90truemm, bb= 70 0 500 350, angle=0,clip=}}}
\vskip2mm
\captionb{4}{The H$\beta$ (left) and H$\delta$ (right) lines in the
            spectrum of KS\,Per. Solid line -- the observed spectrum,
	    triangles -- calculated lines with H/He=10$^{-5}$,
            filled circles -- H/He=$2{\cdot}10^{-5}$,
	    diamonds -- H/He=$5{\cdot}10^{-5}$. }
\vskip2mm
\end{figure}

\clearpage
\newpage
\begin{wrapfigure}{5}[0pt]{151mm}
\centerline{\hspace{0cm}\psfig{figure=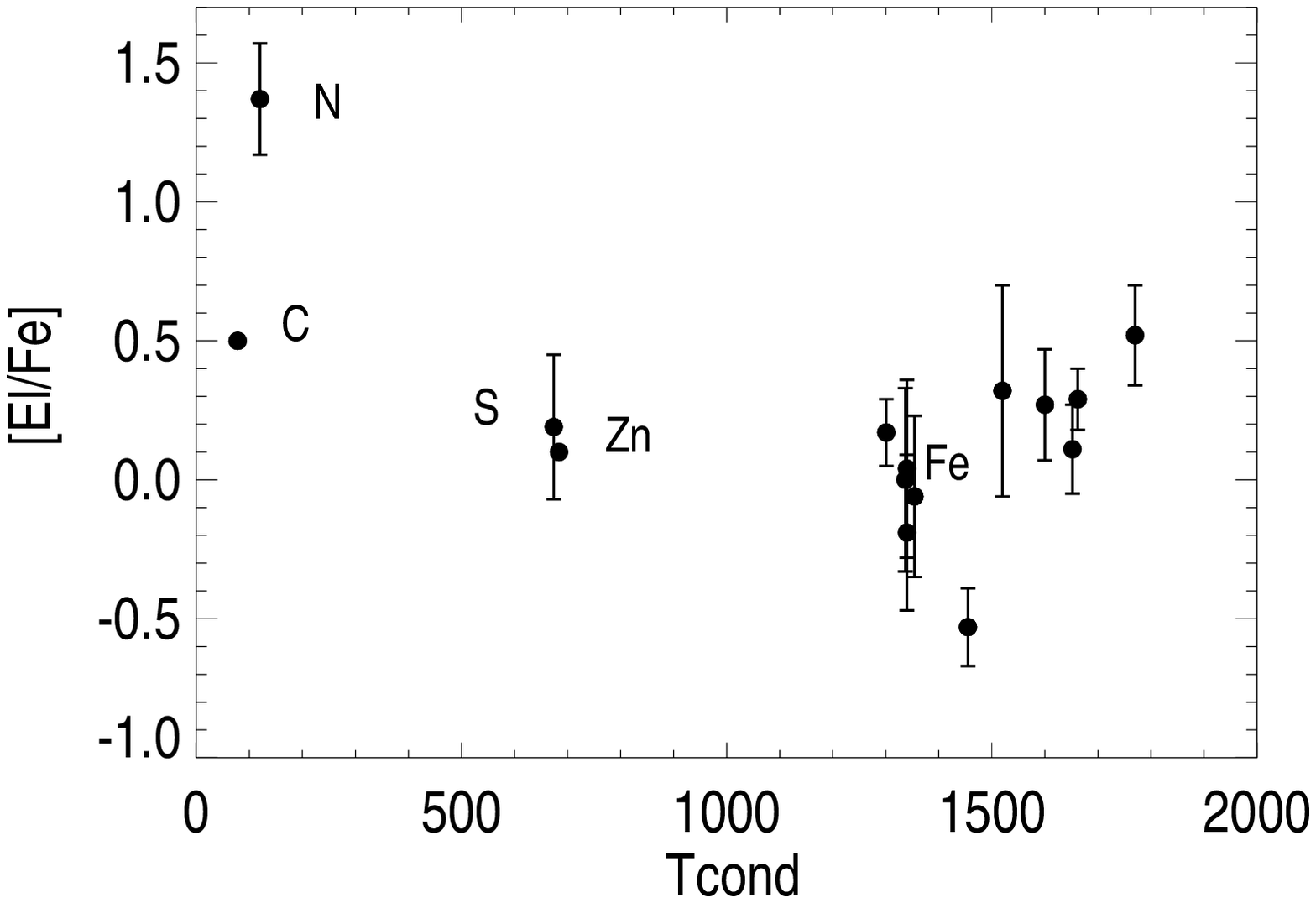,width=150truemm,angle=0,clip=}}
\captionb{5}{The abundances of the elements in the atmospere of KS\,Per
            versus their condensation temperature.}
\end{wrapfigure}

\clearpage
\newpage
\begin{wrapfigure}{c}[0pt]{5.5cm}
\vbox{
\begin{tabular}{lr}
\multicolumn{2}{c}{\parbox{6cm}{\baselineskip=8pt
  ~{\bf Table\,1.}{\ Basic parameters of KS\,Per (SIMBAD
  database).}}}\\
\tablerule
KS\,Per&=HD\,30353\\
\tablerule
 $\alpha_{\rm 2000}$&04 48 53.35\\
 $\delta_{\rm 2000}$&43 16 32.10\\
\tablerule
Galactic~~~~~ $l$&~161.73\\
coordinates~ $b$&~--01.01\\
\tablerule
Mean magnitude $B$&~~8.14\\
               $V$&~~7.76\\
\tablerule
Spectral type &~~A5Iap\\
\tablerule
IRAS fluxes $f_{12}$&~~~~1.58\\
(Jy)        $f_{25}$&~~~~0.52\\
            $f_{60}$&~~~~0.40\\
            $f_{100}$&~~~~1.35\\
\tablerule
\end{tabular}
}
\end{wrapfigure}

\clearpage
\newpage

\begin{table}[!th]
\begin{center}
\vbox{
\begin{tabular}{lllr l}
\multicolumn{5}{c}{\parbox{110mm}{\baselineskip=9pt
{\bf Table\,2.}{\ The chemical composition of KS\,Per.
The abundances are normalised so that $\log \Sigma \mu_i\varepsilon(i)$\,=\,12.15.
In the Remarks the number of used lines and the source of oscillator
strengths are indicated.}}}\\[+4pt]
\tablerule
 & Sun$^1$ &\multicolumn{2}{c}{KS\,Per} &  \\
\cline{3-4}
 El. & $\log\varepsilon$ & $\log\varepsilon$ & [El/Fe] & Remarks \\
\tablerule
H  & 12.00 & 7.0 & & \\
He & 10.93 & $11.54\pm0.23$ &  & \\
C  & 8.39  & 8.1            &      & Input abundance \\
N  & 7.78  & $8.36\pm0.20$  &  1.4 & 11 N\,I, 5 N\,II, WFD$^2$\\
O  & 8.66  & $7.72\pm0.17$  & --0.2 & 3 O\,I, WFD\\[+5pt]
Ne & 7.84  & $7.66\pm0.14$  &  0.6 & 8 Ne\,I, K$^3$\\
Mg & 7.53  & $6.55\pm0.28$  & --0.2& 4 Mg\,I, T$^4$, 7 Mg\,II, T, K\\
Si & 7.51  & $6.76\pm0.32$  &  0.0 & 16 Si\,II, K, A$^5$\\
S  & 7.14  & $6.54\pm0.26$  &  0.2 & 7 S\,II, K\\
Sc & 3.05  & $2.37\pm0.16$  &  0.1 & 3 Sc\,II, T \\[+5pt]
Ti & 4.90  & $4.38\pm0.20$  &  0.3& 19 Ti\,II, T\\
V  & 4.00  & $2.68\pm0.14$  & --0.5& 3 V\,II, T\\
Cr & 5.64  & $5.02\pm0.12$  &   0.2& 27 Cr\,II, T\\
Fe & 7.45  & $6.66\pm0.33$  &      & 6 Fe\,I, 33 Fe\,II, T\\
Ni & 6.23  & $5.38\pm0.29$  & --0.1& 3 Ni\,II, T\\[+5pt]
Sr & 2.92  & $2.23\pm0.09$  &   0.1& 2 Sr\,II, T\\
Y  & 2.24  & $1.74\pm0.11$  &   0.3& 2 Y\,II, T\\
Zr & 2.59  & $2.32\pm0.18$  &   0.5& 5 Zr\,II, T\\
Ba & 2.17  & $1.70\pm0.38$  &   0.3& 3 Ba\,II, T\\     
\tablerule
\end{tabular}
}
\end{center}
\centerline{\parbox{110mm}{ \small\baselineskip=9pt
$^1$ ~Asplund et al. (2005), relative to $\log \varepsilon({\rm
H})$,}}
\centerline{\parbox{110mm}{\small\baselineskip=9pt
$^2$ ~Wiese et al. (1996),}}
\centerline{\parbox{110mm}{ \small\baselineskip=9pt
$^3$ ~Thevenin (1989, 1990),}}
\centerline{\parbox{110mm}{ \small\baselineskip=9pt
$^4$ ~van Hoof (1999),}}
\centerline{\parbox{110mm}{ \small\baselineskip=9pt
$^5$ ~Artru et al. (1981).}}
\end{table}

\clearpage
\newpage
\begin{wrapfigure}{l}[0pt]{95mm}
\vbox{
\begin{tabular}{lrrr}
\multicolumn{4}{c}{\parbox{85mm}{\baselineskip=9pt
{\bf Table\,3.}{\ The chemical abundances pattern of KS\,Per
          in comparison to that of the extreme helium star
          LS\,IV$-14^{\rm o} 109$ (Pandey et al., 2001).}}}\\[+4pt]
\tablerule
 & \multicolumn{3}{c}{$\log\varepsilon$ } \\
\cline{2-4}
 El. & Sun &KS\,Per&  LS\,IV$-14^{\rm o} 109$ \\
     &     &9500\,K, 2.0&  9500\,K, 0.9   \\
\tablerule
H  & 12.00 & 7.0   &  6.2  \\
He & 10.93 &11.54  & 11.54 \\
C  & 8.39  & 8.1   &  9.4  \\
N  & 7.78  & 8.36  &  8.6  \\
O  & 8.66  & 7.72  &  8.5  \\
Ne & 7.84  & 7.67  &  9.4  \\
Mg & 7.53  & 6.55  &  6.9  \\
Si & 7.51  & 6.76  &  7.8  \\
S  & 7.14  & 6.54  &  7.6  \\
Sc & 3.05  & 2.37  &  3.3  \\
Ti & 4.90  & 4.38  &  4.3  \\
Cr & 5.64  & 5.02  &  5.1  \\
Fe & 7.45  & 6.66  &  6.8  \\
Ni & 6.23  & 5.38  &  6.6  \\
Sr & 2.92  & 2.23  &  2.6  \\
Y  & 2.24  & 1.74  &  1.9  \\
Zr & 2.59  & 2.32  &  1.9  \\
Ba & 2.17  & 1.70  &  1.7  \\
\tablerule
\end{tabular}
}
\end{wrapfigure}

\clearpage
\newpage

\begin{table}[h]
\begin{center}
\vbox{\footnotesize\tabcolsep=4.7pt
\begin{tabular}{lcrrrc@{\hhuad}|@{\hhuad}lcrrrc}
\multicolumn{12}{c}{\parbox{122mm}{\baselineskip=9pt
{\bf Table\,4.}{\ Atomic data (wavelengths [nm], lower level
excitation potentials $\epsilon_i$ [eV], oscillator strengths $\log gf$),
equivalent widths [pm] and abundances calculated with the model
parameters $T_{\rm eff}=9500$ K, $\log g=2.0$ and 
$\xi_{\rm t}=9.5$\,km\,s$^{-1}$ for HD\,30353.\lstrut}}}\\
\hline
El. & $\lambda$ & $\epsilon_i$~~ & $\log gf$ & $EW$ &\hstrut
 $\log \varepsilon$& El. & $\lambda$ & $\epsilon_i$~~ & $\log gf$ & $EW$ &  $\log \varepsilon$\\[+3pt]
\hline
He\,I& 400.926& 21.22& -1.45& 45.1& 0.12& Si\,II& 407.678 &  9.84& 1.55& 36. & -4.28 \hstrut\\
He\,I& 402.926& 20.96& -0.47& 60.0& -0.07&Si\,II& 437.697 & 12.84& -0.84& 24.4& -4.28\\ 
He\,I& 414.376& 21.22& -1.20& 47.7&  0.06&Si\,II& 467.328& 12.84& -0.39& 22.7& -4.83\\ 
He\,I& 492.193& 21.22& -0.44& 52.3& -0.31&Si\,II& 469.314& 12.15& -1.74&  7.0& -4.78\\
He\,I& 504.774& 21.22& -1.60& 44.2&  0.39&Si\,II& 477.430& 12.84& -1.84&  7.9& -4.29\\
&&&&&&                                    Si\,II& 484.479& 12.88& -1.36&  6.7& -4.84\\
N\,I & 409.950& 12.01& -1.46& 24.1& -2.80&Si\,II& 546.215& 12.87& -1.02& 14.2& -4.67\\ 
N\,I & 410.995& 10.69& -1.23& 34.5& -3.22&Si\,II& 546.946& 12.88& -0.72& 17.9& -4.75\\
N\,I & 599.943& 11.60& -1.11& 22.6& -3.49&Si\,II& 557.597& 12.88& -1.25& 13.3& -4.48\\ 
N\,I & 600.847& 11.60& -1.41& 25.0& -3.09&Si\,II& 568.886& 14.17&  0.08& 14.7& -5.08\\ 
N\,I & 664.650& 11.75& -1.54& 17.0& -3.22&Si\,II& 570.138& 14.17& -0.10& 13.0& -5.02\\ 
N\,I & 665.346& 11.75& -1.14& 26.3& -3.26&Si\,II& 580.050& 14.49& -0.17& 16.3& -4.58\\
N\,I & 561.654& 11.71& -1.32& 22.2& -3.23&Si\,II& 597.893& 10.07& -0.04& 44.2& -5.50\\
N\,I & 562.320& 11.71& -1.60& 16.4& -3.19&Si\,II& 634.711&  8.12&  0.18& 88.5& -4.70\\
N\,I & 581.650& 11.83& -1.97& 12.5& -2.93&Si\,II& 637.137&  8.12& -0.12& 80.4& -4.75\\
N\,I & 664.500& 11.71& -0.91& 29.5& -3.38&Si\,II& 667.190& 14.49&  0.52& 23.9& -4.82\\    
N\,I & 493.512& 10.69& -1.89& 19.2& -3.29&  &&&&&\\
&&&&&&                                    S\,II& 426.776& 16.10&  0.30& 16.2& -4.83\\
N\,II& 460.715& 18.48& -0.51&  8.3& -3.32&S\,II& 446.358& 15.94&  0.22&  9.5& -5.24\\
N\,II& 461.387& 18.48& -0.67&  7.6& -3.22&S\,II& 448.343& 15.90& -0.06& 13.2& -4.71\\
N\,II& 464.309& 18.48& -0.36& 10.4& -3.30&S\,II& 448.663& 15.87& -0.46&  7.6& -4.74\\
N\,II& 567.956& 18.48&  0.25& 18.9& -3.13&S\,II& 471.627& 13.62& -0.22& 18.7& -5.23\\
N\,II& 637.962& 18.47& -0.95&  6.5& -2.78&S\,II& 491.720& 14.00& -0.33& 12.0& -5.35\\
&&&&&                                    &S\,II& 543.280& 13.62&  0.16& 31.1& -4.85\\ 
O\,I & 533.073& 10.74& -0.87& 21.0& -4.08&  &&&&&\\
O\,I & 615.817& 10.74& -0.30& 39.6& -3.95&Sc\,II& 424.684&  0.31&  0.36& 47.0& -9.22\\
O\,I & 543.577& 10.74& -1.54& 13.0& -3.77&Sc\,II& 440.040&  0.61& -0.28& 36.2& -8.99\\
&&&&&                                    &Sc\,II& 552.682&  1.77&  0.18& 25.4& -9.30\\
Ne\,I& 503.775& 18.56& -0.82&  4.8& -4.05& &&&&&\\
Ne\,I& 585.249& 16.85& -0.49& 19.9& -3.99&Ti\,II& 402.835&  1.89& -0.89& 46.8& -6.83\\ 
Ne\,I& 588.189& 16.62& -0.77& 20.9& -3.74&Ti\,II& 428.789&  1.08& -1.59& 32.4& -7.50\\
Ne\,I& 597.553& 16.62& -1.27& 10.9& -3.86&Ti\,II& 431.680&  2.05& -1.74& 24.5& -7.05\\
Ne\,I& 614.306& 16.62& -0.10& 30.4& -3.85&Ti\,II& 433.071&  1.17& -2.28& 22.8& -7.11\\
Ne\,I& 616.359& 16.72& -0.62& 21.0& -3.82&Ti\,II& 434.137&  1.12& -2.20& 27.4& -7.07\\
Ne\,I& 626.650& 16.72& -0.30& 26.8& -3.80&Ti\,II& 442.194&  2.06& -1.72& 25.5& -7.05\\
Ne\,I& 640.225& 16.62&  0.33& 35.6& -3.92&Ti\,II& 445.049&  1.08& -1.47& 39.2& -7.37\\
&&&&&                                    &Ti\,II& 447.086&  1.16& -2.21& 30.6& -6.96\\ 
Mg\,I& 470.299&  4.35& -0.55& 13.0& -4.55&Ti\,II& 448.833&  3.12& -0.65& 35.6& -7.08\\
Mg\,I& 516.733&  2.71& -0.75& 30.2& -4.78&Ti\,II& 452.949&  1.57& -1.72& 28.3& -7.27\\         
Mg\,I& 517.270&  2.71& -0.32& 35.6& -5.02&Ti\,II& 458.995&  1.24& -1.64& 43.2& -6.98\\
Mg\,I& 518.362&  2.72& -0.08& 29.4& -5.47&Ti\,II& 470.867&  1.24& -2.38& 19.3& -7.17\\
&&&&&                                    &Ti\,II& 477.998&  2.05& -1.50& 21.6& -7.47\\
Mg\,II& 439.057& 10.00& -0.52& 42.0& -4.96&Ti\,II& 479.854&  1.08& -2.74& 17.1& -7.00\\
Mg\,II& 442.799& 10.00& -1.21& 29.2& -4.96&Ti\,II& 487.401&  3.09& -1.01& 28.5& -7.08\\
Mg\,II& 443.399& 10.00& -0.90& 37.1& -4.86&Ti\,II& 512.916&  1.89& -1.27& 35.0& -7.41\\
Mg\,II& 448.133&  8.86&  0.74& 65.9& -5.44&Ti\,II& 526.215&  1.58& -2.30& 20.9& -7.06\\
Mg\,II& 473.959& 11.57& -0.42& 31.1& -4.85&Ti\,II& 533.679&  1.58& -1.65& 30.9& -7.40\\
Mg\,II& 485.110& 11.63& -0.42& 34.5& -4.65&Ti\,II& 538.103&  1.57& -2.05& 26.2& -7.16\\
Mg\,II& 654.594& 11.63&  0.41& 41.1& -5.30& &&&&&\\
\hline
\end{tabular}
}
\end{center}
\end{table}

\clearpage
\newpage

\begin{table}[!ht]
\begin{center}
\vbox{\footnotesize\tabcolsep=4.7pt
\vskip3mm
\begin{tabular}{lcrrrc@{\hhuad}|@{\hhuad}lcrrrc}
\multicolumn{12}{l}{\parbox{122mm}{\baselineskip=9pt
{\bf Table\,4.}{\ Continued.\lstrut}}}\\[+4pt]
\hline
El. & $\lambda$ & $\epsilon_i$~~ & $\log gf$ & $EW$ &\lstrut\hstrut
 $\log \varepsilon$&El. & $\lambda$ & $\epsilon_i$~~ & $\log gf$ & $EW$ & $\log \varepsilon$ \\[+4pt]
\hline
\noalign{\vspace{1mm}}
V\,II & 403.561& 1.79& -0.12& 30.5& -8.86&Fe\,II& 466.371&  2.89& -4.06& 29.9& -4.87\\
V\,II & 430.110& 4.02&  0.68& 14.5& -9.00&Fe\,II& 466.675&  2.83& -3.53& 45.5& -4.71\\
V\,II & 533.266& 2.27& -1.06&  8.6& -8.72&Fe\,II& 512.035&  2.83& -4.37& 25.0& -4.87\\
&&&&&&                                     Fe\,II& 513.267&  2.81& -4.23& 21.0& -5.16\\
Cr\,II& 413.241&  3.76& -2.35& 29.4& -6.41&Fe\,II& 523.463&  3.22& -2.31& 50.7& -5.61\\
Cr\,II& 424.238&  3.87& -1.11& 50.3& -6.52&Fe\,II& 525.693&  2.89& -4.32& 29.7& -4.72\\
Cr\,II& 425.263&  3.86& -2.00& 34.7& -6.48&Fe\,II& 528.411&  2.89& -3.31& 52.1& -4.76\\
Cr\,II& 427.556&  3.86& -1.45& 41.5& -6.71&Fe\,II& 532.556&  3.22& -3.38& 34.3& -5.28\\
Cr\,II& 455.499&  4.07& -1.53& 45.5& -6.41& Fe\,II& 533.773&  3.23& -3.98& 27.2& -4.94\\
Cr\,II& 455.865&  4.07& -0.67& 61.0& -6.34& Fe\,II& 536.287&  3.20& -2.80& 41.7& -5.57\\
Cr\,II& 459.206&  4.07& -1.51& 47.2& -6.36& Fe\,II& 540.882&  5.95& -2.19& 31.4& -5.02\\
Cr\,II& 461.879&  4.07& -1.21& 48.1& -6.60&Fe\,II& 552.514&  3.27& -4.21& 27.6& -4.71\\
Cr\,II& 463.408&  4.07& -1.25& 51.5& -6.39&Fe\,II& 553.485&  3.24& -2.96& 40.8& -5.46\\
Cr\,II& 481.235&  3.86& -1.99& 36.4& -6.55& Fe\,II& 581.367&  5.57& -2.69& 31.2& -4.79\\
Cr\,II& 482.414&  3.87& -0.94& 57.4& -6.55& Fe\,II& 599.138&  3.15& -3.76& 28.3& -5.25\\
Cr\,II& 483.624&  3.86& -2.18& 39.7& -6.23& Fe\,II& 608.410&  3.20& -3.99& 26.5& -5.06\\
Cr\,II& 486.022&  3.87& -2.21& 35.2& -6.40&Fe\,II& 611.333&  3.22& -4.26& 26.5& -4.77\\
Cr\,II& 486.432&  3.86& -1.50& 51.8& -6.35&Fe\,II& 614.774&  3.89& -2.92& 40.8& -5.19\\
Cr\,II& 487.640&  3.85& -1.68& 42.5& -6.62&Fe\,II& 623.350&  5.48& -2.70& 31.3& -4.87\\
Cr\,II& 488.460&  3.86& -2.24& 30.3& -6.58& Fe\,II& 624.756&  3.89& -2.55& 52.9& -5.09\\
Cr\,II& 524.678&  3.71& -2.55& 23.0& -6.67& Fe\,II& 636.946&  2.89& -4.31& 22.4& -5.09\\
Cr\,II& 524.942&  3.76& -2.64& 24.0& -6.51& Fe\,II& 641.693&  3.89& -2.86& 33.0& -5.57\\
Cr\,II& 527.498&  4.07& -1.55& 44.4& -6.63& &&&&&\\
Cr\,II& 530.843&  4.07& -2.07& 33.5& -6.57&Ni\,II& 401.504 & 4.03 & -2.44& 38.5& -5.98\\
Cr\,II& 531.070&  4.07& -2.34& 29.0& -6.45&Ni\,II& 424.480 & 4.03 & -3.02& 28.2& -6.00\\
Cr\,II& 531.359&  4.07& -1.78& 35.5& -6.77&Ni\,II& 436.210 & 4.03 & -2.34& 32.3& -6.49\\
Cr\,II& 533.776&  4.07& -2.18& 30.0& -6.59& &&&&&\\
Cr\,II& 540.762&  3.83& -2.38& 28.0& -6.61&Sr\,II& 407.772 & 0.00 & 0.17 & 41.1& -9.39\\
Cr\,II& 542.093&  3.76& -2.49& 28.6& -6.52&Sr\,II& 421.554 & 0.00 & -0.38& 34.4& -9.22\\
Cr\,II& 550.863&  4.15& -2.20& 30.0& -6.54& &&&&&\\
Cr\,II& 605.347&  4.73& -2.18& 19.9& -6.62&Y\,II & 417.754&  0.41&  0.20& 25.2& -9.67\\
&&&&&&                                     Y\,II & 430.963&  0.18& -0.42& 11.6& -9.76\\
Fe\,I & 404.582& 1.48& 0.29& 41.2& -4.68& Y\,II & 437.494&  0.41&  0.30& 26.6& -9.75\\
Fe\,I & 406.361& 1.55& 0.00& 33.3& -4.75& &&&&&\\
Fe\,I & 407.175& 1.60& -0.04& 22.8& -5.12&Zr\,II& 414.920&  0.80&  0.08& 22.9& -9.46\\
Fe\,I & 429.925& 2.42& -0.72& 11.4& -4.46&Zr\,II& 420.899&  0.71& -0.51& 20.5& -9.04\\
Fe\,I & 440.476& 1.56& -0.25& 30.9& -4.70&Zr\,II& 421.190&  0.53& -0.65& 19.2& -9.06\\
Fe\,I & 526.955& 0.86& -1.42& 15.6& -4.65&Zr\,II& 435.974&  1.24& -0.25& 12.5& -9.30\\
&&&&&&                                    Zr\,II& 449.697&  0.71& -0.85&  9.3& -9.22\\
Fe\,II& 412.267&  2.58& -3.53& 45.2& -4.66& &&&&&\\
Fe\,II& 412.479&  2.54& -4.26& 32.0& -4.65&Ba\,II& 455.404&  0.00&  0.93& 23.7& -10.36\\
Fe\,II& 427.333&  2.70& -3.31& 34.3& -5.45&Ba\,II& 493.409&  0.00& -0.10& 16.0& -9.68\\
Fe\,II& 436.941&  2.78& -3.82& 43.1& -4.48&Ba\,II& 614.173&  0.70&  0.27& 23.0& -9.48\\
Fe\,II& 448.918&  2.83& -3.15& 34.2& -5.58& &&&&&\\
Fe\,II& 452.023&  2.81& -2.88& 50.1& -5.06& &&&&&\\
Fe\,II& 457.634&  2.84& -3.22& 36.8& -5.41& &&&&&\\
Fe\,II& 458.283&  2.84& -3.41& 46.0& -4.78& &&&&&\\
Fe\,II& 462.052&  2.83& -3.52& 45.0& -4.73& &&&&&\\
Fe\,II& 463.531&  5.95& -1.60& 34.8& -5.37& &&&&&\\
Fe\,II& 464.895&  2.58& -4.62& 21.1& -4.83& &&&&&\\
\hline
\end{tabular}				     
}					     
\end{center}				    
\end{table}
%%%%%%

\end{document}